\input harvmac
\input epsf
%
%
\noblackbox
\newcount\figno
\figno=0
\def\fig#1#2#3{
\par\begingroup\parindent=0pt\leftskip=1cm\rightskip=1cm\parindent=0pt
\baselineskip=11pt
\global\advance\figno by 1
\midinsert
\epsfxsize=#3
\centerline{\epsfbox{#2}}
\vskip -21pt
{\bf Fig.\ \the\figno: } #1\par
\endinsert\endgroup\par
}
\def\figlabel#1{\xdef#1{\the\figno}}
\def\encadremath#1{\vbox{\hrule\hbox{\vrule\kern8pt\vbox{\kern8pt
\hbox{$\displaystyle #1$}\kern8pt}
\kern8pt\vrule}\hrule}}

\def\frac#1#2{{#1 \over #2}}

\def\p{\partial}
\def\semi{\subset\kern-1em\times\;}
\def\bar#1{\overline{#1}}
\def\sqr#1#2{{\vcenter{\vbox{\hrule height.#2pt
\hbox{\vrule width.#2pt height#1pt \kern#1pt \vrule width.#2pt}
\hrule height.#2pt}}}}

\def\p{\partial}
\def\pb{\bar{\partial}}

\def\h{{1 \over 2}}

\def\p{\partial}
\def\rt{\tilde{r} }

\def\cL{ {\cal L} }
\def\zb{\bar{z}}

\def\Tr{{\rm Tr}}

\def\cF{{\cal F}}
\def\rt{\tilde{r} }
\def\tg{\tilde g}
\def\tB{\tilde B}
\def\tC{\tilde C}
\def\vt{\vartheta}

\lref\myers{ R.~C.~Myers,
``Dielectric-branes,''
JHEP {\bf 9912}, 022 (1999)
[arXiv:hep-th/9910053].
 }
\lref\supertube{
D.~Mateos and P.~K.~Townsend,
``Supertubes,''
Phys.\ Rev.\ Lett.\  {\bf 87}, 011602 (2001)
[arXiv:hep-th/0103030].
}

\lref\mathurlunin{
O.~Lunin and S.~D.~Mathur,
``Metric of the multiply wound rotating string,''
Nucl.\ Phys.\ B {\bf 610}, 49 (2001)
[arXiv:hep-th/0105136].
}

\lref\LuninJY{
O.~Lunin and S.~D.~Mathur,
``AdS/CFT duality and the black hole information paradox,''
Nucl.\ Phys.\ B {\bf 623}, 342 (2002)
[arXiv:hep-th/0109154].
}

\lref\mathurstretch{
O.~Lunin and S.~D.~Mathur,
 ``Statistical interpretation of Bekenstein entropy for systems with a
stretched horizon,''
Phys.\ Rev.\ Lett.\  {\bf 88}, 211303 (2002)
[arXiv:hep-th/0202072].
}

\lref\LuninBJ{
O.~Lunin, S.~D.~Mathur and A.~Saxena,
``What is the gravity dual of a chiral primary?,''
Nucl.\ Phys.\ B {\bf 655}, 185 (2003)
[arXiv:hep-th/0211292].
}

\lref\LuninIZ{
O.~Lunin, J.~Maldacena and L.~Maoz,
``Gravity solutions for the D1-D5 system with angular momentum,''
arXiv:hep-th/0212210.
}

\lref\mathur{
S.~D.~Mathur, A.~Saxena and Y.~K.~Srivastava,
``Constructing 'hair' for the three charge hole,''
arXiv:hep-th/0311092.
}

\lref\MathurSV{
S.~D.~Mathur,
``Where are the states of a black hole?,''
arXiv:hep-th/0401115.
}

\lref\ifd{
I.~Bena,
``The polarization of F1 strings into D2 branes: 'Aut Caesar aut nihil',''
Phys.\ Rev.\ D {\bf 67}, 026004 (2003)
[arXiv:hep-th/0111156].
}
\lref\nbi{
A.~A.~Tseytlin,
``On non-abelian generalisation of the Born-Infeld action in string  theory,''
Nucl.\ Phys.\ B {\bf 501}, 41 (1997)
[arXiv:hep-th/9701125].
}
\lref\HerdeiroAP{
C.~A.~R.~Herdeiro,
``Special properties of five dimensional BPS rotating black holes,''
Nucl.\ Phys.\ B {\bf 582}, 363 (2000)
[arXiv:hep-th/0003063].
}
\lref\unp{W.~I.~Taylor, unpublished}
\lref\wati{
W.~I.~Taylor,
``Adhering 0-branes to 6-branes and 8-branes,''
Nucl.\ Phys.\ B {\bf 508}, 122 (1997)
[arXiv:hep-th/9705116].
}
\lref\tvr{
W.~I.~Taylor and M.~Van Raamsdonk,
``Multiple Dp-branes in weak background fields,''
Nucl.\ Phys.\ B {\bf 573}, 703 (2000)
[arXiv:hep-th/9910052].
W.~I.~Taylor and M.~Van Raamsdonk,
``Multiple D0-branes in weakly curved backgrounds,''
Nucl.\ Phys.\ B {\bf 558}, 63 (1999)
[arXiv:hep-th/9904095].
}
\lref\bak{
D.~Bak and K.~M.~Lee,
``Noncommutative supersymmetric tubes,''
Phys.\ Lett.\ B {\bf 509}, 168 (2001)
[arXiv:hep-th/0103148].
}
\lref\emparan{
R.~Emparan, D.~Mateos and P.~K.~Townsend,
``Supergravity supertubes,''
JHEP {\bf 0107}, 011 (2001)
[arXiv:hep-th/0106012].
}
\lref\bmpv{
J.~C.~Breckenridge, R.~C.~Myers, A.~W.~Peet and C.~Vafa,
``D-branes and spinning black holes,''
Phys.\ Lett.\ B {\bf 391}, 93 (1997)
[arXiv:hep-th/9602065].
}

\lref\HorowitzNW{ G.~T.~Horowitz and J.~Polchinski,
``A correspondence principle for black holes and strings,''
Phys.\ Rev.\ D {\bf 55}, 6189 (1997) [arXiv:hep-th/9612146].
}

\lref\BakKQ{
D.~Bak and K.~M.~Lee,
``Noncommutative supersymmetric tubes,''
Phys.\ Lett.\ B {\bf 509}, 168 (2001)
[arXiv:hep-th/0103148].
}

\lref\EmparanUX{
R.~Emparan, D.~Mateos and P.~K.~Townsend,
``Supergravity supertubes,''
JHEP {\bf 0107}, 011 (2001)
[arXiv:hep-th/0106012].
}

\lref\BenaWP{
I.~Bena,
``The polarization of F1 strings into D2 branes: 'Aut Caesar aut nihil',''
Phys.\ Rev.\ D {\bf 67}, 026004 (2003)
[arXiv:hep-th/0111156].
}

\lref\MateosPR{
D.~Mateos, S.~Ng and P.~K.~Townsend,
``Tachyons, supertubes and brane/anti-brane systems,''
JHEP {\bf 0203}, 016 (2002)
[arXiv:hep-th/0112054].
}

\lref\Buscher{T.~H.~Buscher,
Phys.\ Lett.\ B {\bf 159}, 127 (1985),\ B {\bf 194}, 59 (1987),\ B {\bf 201}, 466 (1988),
}

\lref\MeessenQM{ P.~Meessen and T.~Ortin, ``An Sl(2,Z) multiplet of nine-dimensional type II
supergravity theories,'' Nucl.\
Phys.\ B {\bf 541}, 195 (1999) [arXiv:hep-th/9806120].
}

\lref\don{
G.~T.~Horowitz and D.~Marolf,
``Counting states of black strings with traveling waves. II,''
Phys.\ Rev.\ D {\bf 55}, 846 (1997)
[arXiv:hep-th/9606113].
}

\lref\aspects{
M.~Kruczenski, R.~C.~Myers, A.~W.~Peet and D.~J.~Winters,
``Aspects of supertubes,''
JHEP {\bf 0205}, 017 (2002)
[arXiv:hep-th/0204103].
}

\lref\CallanHN{ C.~G.~Callan, J.~M.~Maldacena and A.~W.~Peet, ``Extremal Black Holes As
Fundamental Strings,'' Nucl.\ Phys.\ B
{\bf 475}, 645 (1996) [arXiv:hep-th/9510134].
}

\lref\DabholkarNC{ A.~Dabholkar, J.~P.~Gauntlett, J.~A.~Harvey and D.~Waldram, ``Strings as
Solitons \& Black Holes as Strings,''
Nucl.\ Phys.\ B {\bf 474}, 85 (1996) [arXiv:hep-th/9511053].
}

\lref\CveticXH{ M.~Cvetic and F.~Larsen, ``Near horizon geometry of rotating black holes in
five dimensions,'' Nucl.\ Phys.\ B
{\bf 531}, 239 (1998) [arXiv:hep-th/9805097].
}

\lref\EmparanWN{
R.~Emparan and H.~S.~Reall,
``A rotating black ring in five dimensions,''
Phys.\ Rev.\ Lett.\  {\bf 88}, 101101 (2002)
[arXiv:hep-th/0110260].
}

\lref\ReallBH{
H.~S.~Reall,
``Higher dimensional black holes and supersymmetry,''
Phys.\ Rev.\ D {\bf 68}, 024024 (2003)
[arXiv:hep-th/0211290].
}

\lref\ElvangMJ{
H.~Elvang and R.~Emparan,
 ``Black rings, supertubes, and a stringy resolution of black hole
non-uniqueness,''
JHEP {\bf 0311}, 035 (2003)
[arXiv:hep-th/0310008].
}

\lref\EmparanWY{
R.~Emparan,
``Rotating circular strings, and infinite non-uniqueness of black rings,''
arXiv:hep-th/0402149.
}

\Title{
  \vbox{\baselineskip12pt \hbox{hep-th/0402144}
  \hbox{UCLA-03-TEP-29}
  \vskip-.5in}
}{\vbox{
  \centerline{Three Charge Supertubes and Black Hole Hair}
 }}

\centerline{Iosif Bena and Per Kraus}

\bigskip\medskip
\centerline{ \it Department of Physics and Astronomy, UCLA, Los Angeles, CA 90095-1547,
USA}

\medskip
\medskip
\medskip
\medskip
\medskip
\medskip
\baselineskip14pt
\noindent

We construct finite size, supersymmetric, tubular  D-brane configurations with three charges, two
angular momenta and several
brane dipole moments. In type IIA string theory these are tubular configurations with D0, D4 and
F1 charge, as well as  D2, D6 and
NS5 dipole moments. These multi-charge generalizations of  supertubes might have interesting
consequences for the physics of the
D1-D5-P black hole. We study the relation  of the tubes to the spinning BMPV black hole, and
find that they have properties
consistent with describing some of the hair of this black hole.

\Date{February, 2004}
\baselineskip14pt
\newsec{Introduction}

One of the more novel brane configurations considered in recent years is the so-called supertube
\supertube: this is a tubular
D2-brane with worldvolume electric and magnetic fields turned on such that it carries nonzero
values of D0-brane charge,
fundamental string charge, and angular momentum (see \emparan \MateosPR\ for a sampling of
further work). The resulting
configuration is supersymmetric, and remains so even when the cross section of the tube describes
an arbitrary curve, and when
several tubes are considered simultaneously.  The crucial ingredient in the construction of the
supertubes is the presence of a
critical electric field, $2\pi \alpha' E=1$.\foot{To be precise, this is the critical value in the absence
of a magnetic field.}
This leads to the disappearance of the D2-brane from the equations determining the preserved
supersymmetry of the system as well
as its tension; indeed the tension just becomes that of the D0 and F1 constituents.

The original supertube carries two independent conserved charges
(D0 and F1), but it is only natural to consider the
generalization to three charges\foot{Some three charge configurations have
also  been considered in \aspects.}.  This is one of the motivations for
the present work.  Up to U-duality we can take the three
charges to be those of D0-branes, D4-branes, and F1 strings, and this is
the description we will find most convenient. The finite
size of the resulting configuration leads to dipole moments for other
branes.  If we consider the three possible pairings of
charges, and dualize the statement that D0 and F1 charges lead to a D2
dipole moment, we are lead to expect that our configuration
will carry nonzero D2, D6, and NS5 dipole moments.

We will present two independent constructions of the three charge tubes.
In the first we start with a tubular D6-brane, as
described by the Born-Infeld action,  and turn on fluxes so as to induce
the correct lower brane charges.  This is a
straightforward generalization of the original supertube construction in
\supertube.   The second construction is based on the
non-abelian theory of the D4-branes, and involves exciting the transverse
scalars appropriately.  This generalizes the
construction in \ifd.  In both cases our considerations will be
entirely classical, which implies that we cannot see the expected
NS5 dipole moments.  We furthermore expect that upon passing to
the quantum theory our configurations will correspond to
marginally bound states.

Besides their intrinsic interest, supertubes are beginning to play
an important role in black hole physics, based on the work
of  Mathur, Lunin, and collaborators \mathurlunin \LuninBJ \LuninIZ
\LuninJY \mathurstretch \mathur; for a recent review see \MathurSV.
After a chain of dualities, the various configurations of the
two charge supertubes are in one-to-one correspondence with the
supersymmetric ground states of the D1-D5 system (with vanishing momentum).
Furthermore, the corresponding supergravity solutions
have been derived and turn out to
be free of singularities, thus yielding a direct map between classical
geometries and brane
microstates.  In this sense, the supertubes can be thought of as
the hair of the D1-D5 system.

The story becomes even more interesting when we add the third charge,
which corresponds to momentum in the D1-D5 description,
since the system acquires a macroscopically large entropy for large
charges.  It has been conjectured that the supersymmetric
states of the three charge system will continue to be in one-to-one
correspondence with classical geometries, although this has so
far only been checked for a single unit of momentum \mathur.  In analogy
with the above discussion, we would then like to associate these
states with the three charge supertubes which we study in this paper.

Since angular momentum plays an important  role in the supertube construction, we will compare
the properties of our supertubes
with the properties of the rotating D1-D5-P black hole --- the BMPV black hole.  According the
story developed in [9-12] the BMPV
black hole should represent, roughly speaking, the statistical average of the microstates of the
D1-D5-P system with fixed angular
momentum. By comparing the size and angular momentum bounds of our tubes with those of the
BMPV black hole we will see that a
consistent picture emerges. We will also use the tubes as a probe of the BMPV geometry in order
to give support to the idea that
the black hole can be thought of as being made up of tubes.  A nice consistency check  is to see
how one is prevented from
overspinning the black hole (which would result in closed timelike curves) by dropping  high
angular momentum tubes into the
horizon. This provides a rather remarkable example of chronology protection at work.

We should remark that these supertubes are unlike other configurations
used in studying black hole entropy. Usually one computes
the microscopic entropy at weak coupling, where the system is of string
scale in extent, and its Schwarzschild radius even
smaller.  As the gravitational coupling is increased, the Schwarzschild
radius grows, becoming comparable to the size of the brane
configuration at the ``correspondence point'' \HorowitzNW, and larger
thereafter.
There are thus two descriptions of the system: as a microscopic
string theory object for small $g_s$, and as a black hole for large
$g_s$.  One then compares the entropy in the two regimes and finds
an agreement, which is precise if supersymmetry forbids
corrections during the extrapolation.  The supertubes are different.
The size of a  tube is determined by a balance
between the angular momentum of the system and the tension of the
tubular brane. As  the string coupling is increased, the D-brane
tension decreases, and thus the size of the tube grows, much like
one would expect if these configurations directly represent the
black hole microstates even at large $g_s$.

The remainder of this paper is organized as follows.  In section 2 we
present the construction of the tubes from the D6-brane point of view;
this is followed by the construction in terms of D4-branes in section 3.
Connections with black hole physics are studied in section 4.  In section
5 we add some concluding thoughts.  For convenience, we have included
an appendix on the BMPV black hole.
Throughout this paper we will using the word ``supertube'' to denote any of
 the U-duals of the 2 or 3 charge configurations we
construct, even if in the D1-D5-P case these configurations do not
look tubular (they are rotating  helical branes).


\newsec{Construction of the tubes - the D6 brane picture.}


We begin with a single tubular D6-brane, and attempt to turn on
worldvolume fluxes such that we describe a BPS configuration
carrying D4, D0 and F1 charges. Using a single D6-brane also leads
to the presence of D2-brane charges, but we will subsequently
introduce a second D6-brane to cancel this.

The D6-brane is described by the Born-Infeld action
\eqn\za{ S= - T_6  \int\! d^7 \xi ~ \sqrt{ -\det(g_{ab} +\cF_{ab})},}
where $g_{ab}$ is the induced worldvolume metric, $\cF_{ab} = 2\pi F_{ab}$,
and we have set $\alpha'=1$.  The induced  D4-brane
and D0-brane  charge densities  are given by
\eqn\zb{\eqalign{ Q_4& = 2\pi T_6  \cF
\cr Q_0 &=  2\pi T_6 \cF \wedge \cF \wedge \cF~.}}
The F1 charge density is proportional to the canonical momentum conjugate
to the vector potential:
\eqn\zc{Q_1 =  \vec{\pi} = {\p {\cal L} \over \p \dot{\vec{A} }}~. }
Factors of $2\pi$ in \zb\ and \zc\ are deserving of comment.  One direction
of our D6-brane will be an $S^1$, and we have defined the charges
after integrating over the corresponding angular coordinate.  So the D-brane
charges are really charge densities per unit five dimensional area, and the
fundamental string charge is a charge density per unit four dimensional area.
Note also that the charges $Q$ are the ones which appear in the Hamiltonian,
and are related to the number of strings or branes by the corresponding
tensions. These conventions will be convenient later on.

Our construction will essentially follow that of the original
D2-brane supertube, except that we include four extra spatial
dimensions and corresponding fluxes.  We take our D6-brane to have
geometry $R^{1,1} \times S^1 \times T^4$. We take $R^{1,1}$ to
span $x^{0,1}$; $S^1$ to be a circle in the $x^{2,3}$ plane of
radius $r$ and angular coordinate $\theta$; and $T^4$ to span
$x^{6,7,8,9}$.  The configuration carries no net D6-brane charge
due to its tubular shape.  We should note that we are considering a
circular D6-brane just for simplicity, but a general curve in $R^4$
can be considered as in  \MateosPR,  or as  in the probe
analysis we perform in the last section of this paper.

To induce D0-branes we turn on constant values of $\cF_{1 \theta}$, $\cF_{67}, $ and
$\cF_{89}$.    $~\cF_{1 \theta}$ then induces
a density of D4-branes in the $x^{6,7,8,9}$ plane.  To induce F1 charge in the $x^1$ direction
we turn on a constant value of
$\cF_{01}$.  As mentioned above, this single D6-brane configuration also carries D2-brane
charges in the $x^{6,7}$ and $x^{8,9}$
directions, but these will eventually be cancelled by introducing a second D6-brane.

With these fluxes turned on we find
\eqn\zd{ S= - T_6  \int\! d^7 \xi ~ \sqrt{(1-
\cF_{01}^2)r^2+\cF_{1\theta}^2
}\sqrt{(1+\cF_{67}^2)(1+\cF_{89}^2)}.}
By differentiating with respect to $\cF_{01}$ we find
\eqn\ze{ Q_1 =2\pi T_6 {\cF_{01} r^2 \over  \sqrt{(1-
\cF_{01}^2)r^2+\cF_{1\theta}^2
}}\sqrt{(1+\cF_{67}^2)(1+\cF_{89}^2)}.}
The key point to observe now is that if we choose
\eqn\zf{ \cF_{01}=1}
then $r^2$ drops out of the action \zd.  Let us further choose
\eqn\zg{ \cF_{67} = \cF_{89}.}
We can then work out the energy from the canonical Hamiltonian as
\eqn\zh{\eqalign{ H =   \int \! Q_1 \cF_{01}-L &=  \int \! \Big[
Q_1 +2\pi T_6 |\cF_{1\theta}| + 2\pi T_6 |\cF_{1\theta} \cF_{67}
\cF_{89}|\Big] \cr & = \int\! \left[ Q_1 + Q_4 +Q_0 \right]. }}
The final two integrals are over the five noncompact directions of
the D6-brane.
The radius of the system is determined by inverting \ze:
\eqn\zi{ r^2  = {Q_1 \over 2\pi T_6} {\cF_{1\theta}\over 1+ \cF_{67}
\cF_{89}}  = {1 \over (2\pi T_6)^2} {Q_1 Q_4^2 \over Q_0 + Q_4}.}
 $\!\!$ From \zh\ we see that we have saturated the BPS bound, and
so our configuration must solve the equations of motion, as can be
verified directly. Supersymmetry can also be verified precisely as
for the original D2-brane supertube.  The presence of the
electric field, $\cF_{01}=1$, causes the D6-brane to drop out of the
equations determining the tension and the unbroken
supersymmetry.  If we set $Q_0=0$ then \zi\ reduces (with the obvious
relabelings) to the radius formula found for the original
D2-brane supertube \supertube.

As we have noted, the above configuration also carries
nonvanishing D2-brane charge associated with $\cF_{1\theta}
\cF_{67}$ and $\cF_{1\theta} \cF_{89}$.  To remedy this we can
introduce a second D6-brane with flipped signs of $\cF_{67}$ and
$\cF_{89}$ \wati. This simply doubles the D4, D0, and F1 charges, while
cancelling the D2 charge.  The $S^1$ of the second tube need not lie
in the same $x^{2,3}$ plane as the first, and we instead generalize
by taking
it to lie in the $x^{4,5}$ plane.  Even more generally, nothing requires
the second $S^1$ to have the same radius as the first (the only
constraint is the cancellation of the D2-brane charges), and so we
will take it to have radius $\tilde{r}$.

More generally, let us introduce $k$ D6-branes, with fluxes
described by diagonal $k\times k$ matrices.  $\cF_{01}$ is equal
to the unit matrix.  We again set $\cF_{67}=\cF_{89}$, and take
$F_{1\theta}$ to have nonnegative diagonal entries to preclude the
appearance of $\overline{D4}$-branes.  The condition of vanishing
D2-brane charge is given by
\eqn\zj{\Tr~ \cF_{1\theta} \cF_{67} =0.}
Finally, the F1 charge is described by taking $Q_1$ to be an
arbitrary diagonal matrix with nonnegative entries.\foot{Quantum
mechanically,  we should demand that $\Tr~ Q_1$ be an integer
to ensure that the total number of F1 strings is integral.} This
results in a BPS configuration of $k$ D6-branes.  In general, each
D6-brane has a different radius; the radius formula is now given
by \zi\ but with the entries replaced by their corresponding
matrices.  Since our matrices are all diagonal, the Born-Infeld
action is unchanged except for the inclusion of an overall trace.
Similarly,  the energy is given by $H =\int\! \Tr\left[ Q_1 + Q_4
+Q_0 \right]$.

In analogy with the behavior of other branes, if we take the $k$ D6-branes
to sit on top of each other we expect that they can
form a marginally bound state.  In the classical description we should then
demand that the radius matrix \zi\ be proportional to
the unit matrix.  Given a choice of magnetic fluxes, this determines the F1
charge matrix $Q_1$ up to an overall
multiplicative constant which parameterizes the radius of the combined system.


As a special case, consider taking all $k$ D6-branes to be identical modulo the
sign of  $\cF_{67}$ and $\cF_{89}$, so that both $\cF_{1\theta}$
and $\cF_{67} \cF_{89}$ are proportional to the unit matrix.\foot{One
could furthermore choose $\Tr \cF_{67} = \Tr \cF_{67} = 0$ to cancel
the D2 charge, but this does not affect the radius formula. }
Then in terms of the total charges, the radius formula is
\eqn\zia{ r^2    = {1 \over k^2(2\pi T_6)^2} {Q_1^{\rm tot} (Q_4^{\rm
tot})^2 \over Q_0^{\rm tot} + Q_4^{\rm tot}}.}

We observe that
 after fixing the conserved charges and imposing equal radii for
the component tubes, there is still freedom in the values of the fluxes.   These can be partially
parameterized in terms of
various nonconserved ``charges'', such as brane dipole moments. Due to the tubular configuration,
our solution carries nonzero D6,
D4, and D2 dipole moments, proportional to
\eqn\zk{\eqalign{ Q^D_6 & = T_6 r   k \cr
Q^D_4 & =T_6  r \Tr \cF_{67} \cr
Q^D_2 & = T_6 r \Tr \cF_{67} \cF_{89} \equiv T_6  r  k_2 . }}
When the $k$ D6-branes which form the tube are coincident, $k_2$ measures
the local D2 brane charge of the tube. It is also possible to see that
both for a single tube, and for $k$ tubes identical up to the
sign of $\cF_{67}$ and $\cF_{89}$, the dipole moments are related:
\eqn\zl{ {Q_2^D \over Q_6^D} ={k_2 \over k} =  {Q_0 \over Q_4}~. }
Furthermore, if $\cF_{67}$ and $\cF_{89}$ are traceless, this tube has no D2 charge and no D4
dipole moment. More general tubes
are not described by \zl, and need not have zero D4 dipole moment when the D2 charge vanishes.
We should also remark that the D2
dipole moment is an essential ingredient in constructing a supersymmetric three charge tube of
finite size. When this dipole
moment goes to zero, the radius of the tube also becomes zero.

Our tube also carries  angular momentum in the $x^{2,3}$ and $x^{4,5}$
planes  in which the $S^1$ factors lie.  The angular momentum densities
of a configuration with $k$ identical D6 branes in the $x^{2,3}$ plane and
$k'$ identical D6 branes in the $x^{4,5}$ plane are:
\eqn\zm{\eqalign{ J_{23} &= 2\pi r T_{0\theta}
= 2\pi T_6 k \sqrt{(1+\cF_{67}^2)(1+\cF_{89}^2)}r^2, \cr
J_{45} &= 2\pi \tilde{r} T_{0\tilde{\theta}}
= 2\pi T_6 k' \sqrt{(1+\cF_{67}^2)(1+\cF_{89}^2)} \tilde{r}^2.}}
Thus, when one adds D0 brane charge to a F1-D4 supertube, the maximum angular
momentum does not change, even if the radius becomes smaller. For completeness,
we should also mention that the shape of the most generic three
charge tube is an arbitrary curve inside $R^4$. The angular momenta can be obtained rather
straightforwardly from this shape.

\subsec{T-duality to the D1-D5-P system}

A T-duality along $x^1$ transforms our D0-D4-F1 tubes into the more
familiar D1-D5-P  configurations.   This T-duality is implemented
by the replacement $2\pi A^1 \rightarrow X^1$.  The nonzero value of
$\cF_{1\theta}$ before the T-duality translates into a nonzero
$\p_\theta X^1$ after.  This means that the resulting D5-brane
is in the shape of a helix whose axis is parallel to $x^1$.   This
is the same as the observation that the D2-brane supertube T-dualizes
into a helical D1-brane.
Since this helical shape is slightly less convenient to work with than
a tube, we have chosen to emphasize the D0-D4-F1 description instead.

\subsec{More general tubes}

Having constructed three charge supertubes with D6 and D2 dipole moments\foot{The D4 dipole
moment of the configurations described
above can be put to zero without loss of generality, and we will not consider it in this section. In
contrast, the D2 and D6
dipole moments cannot be put to zero.}, it is interesting to explore whether one can say anything
about configurations with more
dipole moments. Before proceeding, it is an instructive exercise to understand the physics behind
the radius formula \zia\ for the
supertube with two dipole moments.

Let us consider two simple (two charge) tubes, one of which is made from $Q_0$ D0 branes,
$Q_1'$ fundamental strings, and $k_2$ D2
branes, and the other from $Q_4$ D4 branes, $Q_1''$ fundamental strings, and $k_6$ D6 branes.
If the radii of the two tubes are
the same, then:

\eqn\ka{(2 \pi T_6)^2 r^2 = {Q_1'' Q_4 \over k_6^2} = {Q_1' Q_0 \over k_2^2}\ .}

Let us furthermore require that  $Q_0/Q_4 = k_2 /k_6$. Then, a
short algebraic manipulation brings us to:

\eqn\la{(2 \pi T_6)^2 r^2 = {(Q_1''+Q_1') Q_4 \over k_6^2 (1+{Q_0 \over Q_4})} =
{(Q_1''+Q_1') Q_0 \over k_2^2 (1+{Q_4 \over
Q_0})} = {Q_1^{\rm tot} Q_0 Q_4 \over {k_2 k_6 (Q_0+Q_4)}} .~}

This formula reproduces \zia, and is moreover duality invariant.
Thus, the three charge supertube
with the property  $Q_0/Q_4 = k_2 /k_6$ \zl\ has the same radius and charges
as the superposition of a D0-F1 and a D4-F1 supertube. Note that the
individual F1 charges of the component tubes ($Q_1'$ and $Q_1''$) need
not be quantized. Only their sum is.

The tubes with three charges and D2 and D6 dipole moments we constructed
can be mapped by a chain of dualities to tubes with D2
and NS5 dipole moments, or to tubes with D6 and NS5 dipole moments. These
tubes can again be thought of as a bound state of two
simple two charge tubes. It is quite natural therefore to expect that the
three charge tube with three dipole moments can be
obtained by putting together three simple tubes. The resulting configuration
is still $1/8$ BPS because each supertube has the
supersymmetries of its components.

Let us take a D2 tube  with charges $Q_1',Q_0'$ and D2 dipole moment  $k_2$; a D6
tube with charges $Q_1'',Q_4'$ and D6 dipole
moment $k_6$; and an NS5 tube  with charges $Q_0'',Q_4''$ and NS5 dipole moment
$k_5$. The condition that the radii be equal is:
\eqn\rsq{(2 \pi T_6)^2 r^2 = {Q_1'' Q_4' \over k_6^2} = {Q_1' Q_0' \over k_2^2} =
{Q_4'' Q_0'' \over k_5^2}~.}
The total charges are
\eqn\tot{Q_1^{\rm tot}=Q_1'+Q_1'', \ \ \ \ Q_0^{\rm tot}=Q_0'+Q_0'',
\ \ \ \ Q_4^{\rm tot}=Q_4'+Q_4'' ~,  }
and the angular momentum of the system is:
\eqn\ltot{2 \pi T_6 J = {Q_1'' Q_4' \over  k_6}
+ {Q_1' Q_0' \over  k_2} + {Q_4'' Q_0'' \over k_5}. }
Thus, given the total charges, dipole moments, and angular momentum,
one has six equations with six unknowns \rsq,\tot,\ltot\
which determine the radius of this multi-charge tube.

\newsec{Construction of the tubes - the D4 brane picture.}

When the radius of the system becomes comparable to the string scale,
rather than describing our configurations by the Born-Infeld
action for the D6-branes, it is more appropriate to seek a description
in terms of lower dimensional branes.  In this section we find a solution
representing the three charge supertube in terms of its constituent
D4-branes; this is parallel to the description of the D2-brane
supertube in terms of D0-branes  (the Matrix Theory description).


We start with a collection of $N_4$ D4-branes, and turn on fields such that
it carries D0 and F1 charge.  Just as in the previous section, we first
present a simple solution which also carries D2-brane charge, and then
modify our solution to cancel this.
  As we'll comment on
later, with a simple relabelling, our solution also yields the
familiar D1-D5-P system.

Let the D4-branes be aligned along $x^{6,7,8,9}$.  As described
below, we will distribute the branes over a distance $ N_4\ell$ in
the $x^1$ direction, thus the D4-brane charge density is
\eqn\ba{ Q_4 = { T_4\over \ell}.}
We turn on the worldvolume field strengths
\eqn\b{ \cF_{67}= \cF_{89} = B~  1_{N_4} .}
The D4-branes thus carry lower brane charge densities
\eqn\c{ Q_2  ={T_4 \over N_4 \ell} ~\Tr ~\cF = {B T_4 \over \ell}, \quad Q_0 ={T_4 \over N_4
\ell}   ~\Tr ~ \cF \wedge \cF = { B^2T_4\over \ell}
.}

To induce $F1$ charge along the $x^1$ direction we turn on the
transverse scalars as  \unp\bak\ifd
\eqn\d{\eqalign{  X^1 &=  \ell( j \delta_{ij}) \cr X^2
&=  \half r(a+a^\dagger) \cr X^3 &=  {
i\over 2} r(a^\dagger-a) \cr X^4 &= 0 \cr X^5 &=0  }}
where
\eqn\e{ a_{ij} = \exp\left(-i{\ell\over 2\pi} t\right) \delta_{i-1,j}.}
The form of $X^1$ implies that we are distributing the $N_4$
D4-branes  separated by a distance $\ell$ in the
$x^1$ direction.   For large $N_4$, up to ``boundary effects''
which are subleading in $N_4$, we have the nonvanishing
``transverse field strengths''
\eqn\f{ \eqalign{ \cF_{02} & = \dot{X^2} = {\ell\over 2\pi} X^3 \cr \cF_{03} &
= \dot{X^3} = -{\ell\over 2\pi} X^2  \cr \cF_{12} & = {i\over 2\pi}[X^1,X^2]
=-{\ell\over 2\pi} X^3 \cr \cF_{13} & ={i\over 2\pi} [X^1,X^3] ={\ell\over 2\pi} X^2. }}
Other commutators vanish:
\eqn\g{ [X^2,X^3]=[X^2,X^4] = \cdots = [X^4,X^5]= [A_6,X^1]\cdots =[A_9,X^5]=0.}
An important property, which we will use below, is that up to
boundary effects,
\eqn\gaa{ (X^2)^2+(X^3)^2 = r^2~ 1_{N_4}. }
Thus $r$ plays the role of the radius.

Commutators of field strengths are vanishing, $[\cF_{\mu\nu},\cF_{\mu' \nu'}]=0$,
and we can therefore use the minimally
non-abelian form of the Born-Infeld action:
\eqn\ga{S=-T_4 ~\Tr\int\! d^5 \xi \, \sqrt{ -\det(\eta_{MN} + \cF_{MN} )} }
where $M$ and $N$ run over all spacetime directions.   The F1
string charge density in the $x^5$ direction is found by
substituting $\cF_{01}\rightarrow \cF_{01} - B_{01}$ and expanding
the action to first order in $B_{01}$ as $S\rightarrow S +\int
\!Q_1 B_{01}$. This yields
\eqn\gb{ Q_1=  {T_4(1+B^2) \ell r^2 \over (2\pi)^2}.}

We can now work out the Hamiltonian as
\eqn\o{H = \int \! \Tr P^i \dot{X}^i - \int \!\cL= \int \! |Q_4 +Q_1 +Q_0|. }
 Thus we have again found a BPS saturating configuration, which implies
that the equations of motion must be satisfied, as can be
verified explicitly.

Our configuration has the independent radial parameter $r$
which  essentially parameterizes the  angular
momentum in the $x^{2,3}$ plane.   By computing the
energy momentum tensor we find the angular momentum to be
\eqn\oa{ J_{23} = {T_4\over 2\pi}(1+B^2)r^2= {2 \pi Q_1 Q_4\over T_4}.}

Alternatively, using \ba, \c, and \gb, we can express the radius in terms
of the charges as
\eqn\rr{ r^2={(2\pi)^2 \over T_4^2} { Q_1 Q_4^2 \over Q_4 +Q_0}.}
\oa\ and \rr\ agree with \zm\ and  \zi\ after
we recall that $ T_4=(2\pi)^2 T_6$. We can see again, that when on adds the third
 charge to the two charge supertube the maximal angular momentum does not change.

To describe the more familiar D1-D5-P system, we merely need to
start with a collection of D5-branes aligned along
$x^{1,6,7,8,9}$, and make the change in notation $X^1 \rightarrow
2\pi A_1$.   This is just an implementation of T-duality along
$x^1$.

We now proceed to add in a second collection of D4-branes to cancel the D2-brane charge
appearing in \c.   This is accomplished by
simply flipping a few signs.  To be a bit more general, we can allow the second set of branes to
expand into  the $X^{4,5}$ plane
with radius $\rt$. The solution is
\eqn\dz{\eqalign{ \cF_{67}&= \cF_{89} = B \sigma_3 \otimes 1_{N_4}\cr
  X^1 &=  1_2 \otimes \ell( j \delta_{ij}) \cr X^2
&= \pmatrix{1&0\cr 0&0} \otimes \half r(a+a^\dagger) \cr X^3 &=\pmatrix{1&0\cr 0&0}
\otimes {
i\over 2} r(a^\dagger-a) \cr X^4 &= \pmatrix{0&0\cr 0&1} \otimes
\half\rt(a+a^\dagger) \cr X^5 &= \pmatrix{0&0\cr 0&1} \otimes {i\over 2}
\rt(a^\dagger-a). }}
The analysis proceeds much as before.  The formulas for the charges are now
\eqn\gbz{\eqalign{Q_4&= {2T_4 \over \ell} \cr Q_0 &= {2 B^2 T_4 \over \ell}\cr
 Q_1&= {2 T_4\over (2\pi)^2}(1+B^2) \ell (r^2 +\rt^2) }}
and the angular momenta are
\eqn\oaz{\eqalign{ J_{23} &= {2T_4\over 2\pi} (1+B^2)r^2
=(2\pi)2 T_6(1+B^2) r^2,\cr
  J_{45} &={2T_4\over 2\pi}(1+B^2) \rt^2=(2\pi)2 T_6(1+B^2) \rt^2~.}}
One can also generalize the above construction by deforming the
tube cross-sections to be ellipsoidal. However, the most general
tube cross-section -- an arbitrary closed line in $R^4$ -- does not
seem to be easily obtainable from the nonabelian approach.

\newsec{ Implications for black hole physics.}

We now shift gears and discuss the relation of our supertubes to the black hole physics based on
the D1-D5-P system. One of our
main goals is to argue for a picture in which the spinning BMPV black hole can be thought of as
being made of supertubes. This
proposal will pass several consistency checks, especially relating to bounds on sizes and angular
momenta.  For example, a
supertube implementation  of chronology protection will prevent us from overspinning the black
hole (which would result in closed
timelike curves.)  In the following, we will be working  in the context of the D1-D5-P system, and
denote the quantized charges as
$N_1$, $N_5$, and $N_p$. We correspondingly U-dualize our previous formulas via the
substitutions $N_0 \rightarrow N_p$, $N_4
\rightarrow N_5$, and $N_1 \rightarrow N_1$, and the same for the $Q$s.

\subsec{ Lightning review of the D1-D5-P system}

First, it is helpful to recall a few facts about the $1+1$ dimensional SCFT describing the D1-D5
system.   We have $N_1 N_5$
hypermultiplets comprising $4 N_1 N_5$ bosons and fermions.  The theory has an $SU(2)_L
\times SU(2)_R$ R-symmetry, which can be
identified with the $SO(4)$ rotation group  in the four dimensions transverse to the branes.  The
leftmoving fermions transform as
$2N_1 N_5$ doublets under $SU(2)_L$, while the leftmoving bosons are neutral.  A single
leftmoving fermion thus has equal
eigenvalues for the $SO(4)$ generators $J_{ij}$:
\eqn\pa{J \equiv J_{23}=J_{45}= \pm \half.}

 We will be interested in BPS states in the Ramond-Ramond (RR)
sector, as these are the states relevant to the study of BPS black holes.
States preserving 8 supercharges are the RR vacua with $N_p=0$.
These states carry angular momenta due to the fermion zero modes;
by aligning the zero modes in different ways we get states with
\eqn\pb{ -N_1N_5 \leq J \leq N_1N_5~.}
The  entropy of such states is most easily computed by dualizing to the
F1-P system, and yields
\eqn\pbb{ S = 2\pi \sqrt{N_1N_5 - |J|}, \quad {\rm for} \quad
 |J| \sim N_1N_5.}

States preserving 4 supercharges are obtained by considering purely leftmoving (or rightmoving)
excitations.  The maximal $J$ is
obtained by distributing the $N_p$ units of momentum among as many carriers as possible. This
is achieved by filling up the Fermi
sea.\foot{For our purposes we can think of the CFT as a free theory.}  The counting is simple for
$N_p \gg N_1N_5$, in which case
we  fill up the Fermi sea to the highest harmonic $n_F = \sqrt{{N_p \over N_1 N_5}}$.  The
angular momentum then obeys
\eqn\pc{ -N_1N_5N_p \leq J^2 \leq N_1N_5N_p,}
which is much greater than \pb\ given our assumption about the magnitude
of $N_p$.  For $N_1N_5 N_p \gg J^2$, the entropy was argued in \bmpv\
to be
\eqn\pd{S=2\pi \sqrt{N_1N_5 N_p - J^2}.}

On the supergravity side, the entropy formula \pbb\ was given a stretched horizon type
interpretation in \mathurstretch, while
\pd\ is equal to the Bekenstein-Hawking entropy of the rotating BMPV black hole
\bmpv\CveticXH.  The BMPV solutions indeed obey
the bound \pc; overspinning results in closed timelike curves.

\subsec{ Comparison with supertubes}

Turning now to our supertubes, we note that their angular momenta are not restricted to be equal.
For example, we can choose a
closed curve such that the supertube cross-section lies in the $23$ plane, in which case $J_{23}
\neq 0 $ and $J_{45} =0$.  The
bound on the angular momentum of the tubes coincides with the $N_p=0$ bound \pb. A single
circular tube with $Q_0=0$ saturates
this bound:
\eqn\pe{ J_{23} = {2\pi Q_1 Q_4 \over  T_4} = N_1 N_4\rightarrow N_1 N_5}
while circular tubes with D6 dipole moment $k$ have
\eqn\pf{ J = {2\pi Q_1 Q_4 \over  k T_4} ={ N_1 N_4\over k} \rightarrow {N_1 N_5\over k}~.}
By appropriately changing the shape and orientation of the tube cross
section, we can span the entire range in \pb.

Next consider the three charge supertubes with $r=\rt$.
The angular momentum is still given by \pf, which we write as
$k= N_1 N_5/ J$.  The radius is obtained from \zia\ as:
\eqn\pg{
r^2 = {2\pi \over 8k^2  }{N_1 N_4^2 \over T_0N_0 +T_4 N_4}\rightarrow
{2\pi \over 8k^2  }{N_1 N_5^2 \over N_p +T_5 N_5} ={\pi \over 4}{J^2 \over
N_1 (N_p +T_5 N_5)} }
%
From \pf\ we note that for sufficiently large $N$ we can easily exceed
the black hole angular momentum bound in \pc.  We should also compare
the size of the tubes with the size of the black hole.  For simplicity,
we consider the case of equal charges: $N_5=N_1=N_p$ and $g_s \ll 1$,
yielding
\eqn\pii{ r_{\rm tube}^2 \sim g_s {J^2 \over N^2}~.}
On the other hand, one can use  \bmpv\ to compute the proper length of
the circumference of the horizon (as measured at one of the
equator circles) to be
\eqn\pj{r_{\rm hole}^2 \sim  g_s{N^3 - J^2 \over N^2}~.}
For small $J$ we have $r_{\rm tube} < r_{\rm hole}$, and so we can
consider the tube to fit  inside the horizon.   As $J$ is increased
the tube eventually becomes larger than the horizon, and for $J^2 > N^3$
the black hole ceases to exist.  In fact, since the crossover point is
also $J^2 \sim N^3$, the black hole is essentially always larger
than the tube in the region of parameter
space where both can exist.

It has been proposed \LuninJY-\MathurSV\ that black hole entropy can
be accounted for by the multiplicity of possible
configurations inside the horizon, all of which appear essentially
identical outside the horizon. In this spirit, we can imagine
writing down supergravity solutions for each of our tube configurations.
It is then an important consistency check that the tubes
indeed lie inside the would-be horizon radius, since otherwise the
individual geometries would be distinguishable even outside the
horizon. Although our tubes can become very large, they are almost
never larger than the horizon radius of the corresponding
black hole, and so it is consistent to regard them as comprising part of the
hair of the black hole.

\subsec{AdS - CFT interpretation}

 It is
interesting to see what happens if we try to give an AdS/CFT
interpretation to the supertubes with $J^2 > N^3$.  Recall that in the
CFT of the D1-D5 system, for $N_p \gg N_1 N_5$,  we have the strict bound \pc.
On the other hand, it would seem that we could violate this bound in the
bulk by placing one of our $J^2 > N^3$ tubes in AdS$_3 \times S^3 \times T^4$.
 However, this does not happen, for reasons analogous to the familiar
giant graviton phenomenon.
 In the near
horizon metric of the D1-D5 system, the $S^3$ has a size
\eqn\aea{r_{S^3}^2 = g_s (N_1 N_5)^{1/2} \sim  g_s N.}
In order to fit the tube in the near horizon region we need
$r_{\rm tube} < r_{S^3}$,  which from \pii\ requires $J^2 <N^3$.
So we see that tubes with $J^2 >N^3$ are too large to fit in the
near horizon region.

For  $N_p < N_1 N_5$ the picture that emerges is even more interesting. In that case, there are
many field theory states whose
angular momentum (bounded above by \pb) is larger than the black hole angular momentum
(bounded above by \pc). These states cannot
be described by the black hole alone. Instead, as it has been argued in \LuninBJ\ for the zero
momentum case, these states should
be dual to supertube configurations. Thus, some of the
 states of the field theory can be mapped
to one black hole, whose entropy gives the multiplicity of these states. Other states
however are dual to spacetimes containing supertubes.

It is quite obvious that this distinction is arbitrary, as there is nothing
special about the bound \pc\ in the regime $N_p < N_1 N_5$. It is therefore likely
that all the states of the field theory are dual to supertube configurations.
It is a distinct possibility that in some regime of parameters these
tubes would be rather small and therefore not describable in supergravity.
However, in other regimes the tubes are supergravity objects,
and the possibility of using them to account for the black hole entropy
is very interesting.

If we assume that for a certain angular momentum $J^2 > N_1 N_5 N_p$
each field theory state can be mapped to a supertube
geometry, then we can ask if there exists a ``black object'' which represents
the statistical average of these states.   There are
two possibilities for what this object can be. The first is to have a black
hole together with a supertube at a certain distance
away from the horizon. The other possibility is to have a BPS black ring
solution with $J^2 > N_1 N_5 N_p$, whose hair is given by these supertubes.
Several attempts at constructing nonextremal  and extremal black rings which
carry three charges and several dipole charges have appeared in the
literature \ElvangMJ, but the solutions found have pathologies.
We should also note that a BPS black ring solution would be the first
of its kind, since other known black
rings (\EmparanWN,\EmparanWY) are nonextremal. It would be interesting to
find if such a BPS black ring exists, and if its entropy matches the
multiplicity of the corresponding CFT states.

\subsec{Constructing generalizations of the BMPV black hole}

We noted above that the BMPV black hole obeys $J_{23}=J_{45}$,
while our tubes obey no such restriction.  On the other hand, we
expect our tubes to coexist with the BMPV black hole in a BPS fashion.
This suggests that we can form a more general BPS black
hole by throwing in a tube through the horizon of the BMPV black hole.

We should note that since the D1-D5-P supertube is not homogeneous
along the P direction, it does not directly descend to a five
dimensional configuration. Hence, the resulting black hole is not
properly a  black hole of minimal five dimensional supergravity
\foot{This is also expected from the fact that the only BPS black
hole of this supergravity
is the BMPV black hole \ReallBH.}, but a
six dimensional black string. In fact, some solutions of this
type have been proposed and discussed in \don\ (based on
\CallanHN\DabholkarNC) as gyration waves along the six dimensional
lift of the BMPV black hole. It might be possible to argue that some
of the ten dimensional type IIA black brane configurations we
will be discussing should be dual to smeared gyration waves along
the string.

With this motivation in mind, we now turn to a discussion of the tube treated
as a probe of the BMPV black hole to see if we can
indeed perform this ``experiment''.  We will also discuss the possibility of
creating black holes by putting together three charge
tubes. As before, for convenience we will perform our probe computations
in the D0-D4-F1 picture, but we will sometimes rewrite
our results after the duality which interchanges
 $N_0 \rightarrow N_p$, $N_4 \rightarrow N_5$, and
$N_1 \rightarrow N_1$.

As a warmup we first consider the simpler setup of a D2-brane
supertube probing a D4-brane background. In fact, we will see that this
actually reproduces the results of the more complicated setup.
 If we can slowly contract the tube
down to $r=0$ it will indicate that a bound state can form; up to
T-duality this bound state is a D1-D5-P black hole, and we can use
this experiment to understand the bound  \pc.

We use the string frame metric for a collection of D4-branes
aligned along $x^{0,6,7,8,9}$ and smeared along $x^5$:
\eqn\qa{\eqalign{ ds^2 &= Z^{-1/2}dx_\parallel^2+Z^{1/2}dx_\perp^2 \cr e^{2\phi} &=
Z^{-1/2} \cr Z& =1+{2\pi g_s N_4 \over r^2} =
1+{N_4 \over 2\pi T_2}{1\over r^2}.}}
In the last line ${1 \over r^3} \rightarrow {2 \over r^2}$ due to the
smearing, and we wrote the result in terms of $T_2$ for later convenience.
Note that $N_4$ is the number of D4-branes per unit length in
$x^5$.     Our probe is a D2-brane supertube, with axis parallel
to $x^5$,  and corresponding $S^1$ in the $x^{2,3}$ plane. The
worldvolume fields are $\cF_{05}=1$ and $\cF_{5\theta}$.   The
Born-Infeld action of the tube is
\eqn\qb{ S = -T_2 \int\! d^3 x \, e^{-\phi} \sqrt{-\det
\left(g_{ab} +\cF_{ab}\right)} =-T_2\int \! d^3x \, \cF_{5\theta}
= -\int \! d^2 x \, T_0 N_0. }
As usual, the cancellation of the $Z$ factors follows from the BPS
property.

As in flat spacetime, the radius is determined from the formula
for the fundamental string charge, which in this case is
\eqn\qc{ N_1 = {2\pi T_2 \over N_0} Z r^2.}
Combining \qa\ and \qc\ we arrive at
\eqn\qd{2\pi T_2 r^2 = N_1N_0 - N_4.}
Bound states are described by solutions with $r^2 \leq0$, which
requires $N_1 N_0 \leq N_4$.   On the other hand, the angular
momentum of the supertube is
\eqn\qe{J_{23} = N_0 N_1}
and so we learn that bound states obey the restriction
\eqn\qf{ J_{23} \leq \sqrt{N_0 N_1 N_4}\rightarrow \sqrt{N_1 N_5 N_p}}
which agrees with  \pc.  It is quite remarkable that the BPS black
holes we construct this way, even if they have no angular
momentum in the 45 plane, still obey the BMPV angular momentum
bound \qf. One can also imagine putting together equal amounts of
probe tubes such that the total $J_{23}$ and $J_{45}$ are equal.
Most likely the resulting black hole is the BMPV black hole. The
fact that we can only create BMPV black holes which obey the bound \qf\
and hence have no closed timelike curves, is a remarkable
example of chronology protection at work.

To see how to saturate the bound by bringing in tubes, consider the following
process.  Let the final bound state have charges $N_0$, $N_1$, and $N_4$
(all much larger than $1$).  We can always choose two charges such
that their product is much larger than the third, and so up to duality
we can  take $N_0 N_1\gg N_4$.   Starting from $N_4$ D4-branes,
we consider bringing in $k$ tube probes  each carrying
\eqn\qfa{ \Big( (N_0)_{\rm probe},(N_1)_{\rm probe}\Big) =\left( {N_0 \over k}, {N_1 \over
k}\right) \quad \Rightarrow \quad
J_{\rm probe} = {N_0 N_1 \over k^2}. }
To get as much $J$ as possible we would like $k$ to be as small as
possible, but at the same time we must obey
$(N_0)_{\rm probe}(N_1)_{\rm probe} \leq N_4$ in order to be able
to bring the probe to $r=0$.  Therefore we should choose
$k =\sqrt{ {N_0 N_1 \over N_4}}$, leading to
\eqn\qfb{ J = k J_{\rm probe}= \sqrt{N_0 N_1 N_4}
\rightarrow \sqrt{N_1 N_5 N_p},}
which indeed saturates the bound.

As we have discussed, the black hole we create from supertubes has $J_{45}=0$,
while the BMPV black hole has
$J_{45}=J_{23}$. We can of course introduce nonzero $J_{45}$ by throwing in
supertubes whose $S^1$ factors lie in this plane, and
therefore it is clear that we can obtain BPS configurations with arbitrary
and independent $J_{23}$ and $J_{45}$, such that their sum is within the bound
set by \pc.  It would be interesting to see if the explicit
solution for these configurations can be obtained by smearing gyration
waves on the six dimensional black string \don.

In the CFT description of the D1-D5-P system, angular momentum is
carried by the fermions, the leftmoving species of which have
$J_{23}=J_{45}$, and so the possible angular momentum is more restricted
than our supertube thought experiment would suggest.  Actually, even
for BPS states we are not strictly required to have $J_{23}=J_{45}$,
since angular momentum can also be carried by the rightmoving zeromodes,
which have $J_{23}=-J_{45}=\half$.  This means that we can have
$|J_{23}-J_{45} | \leq 2N_1 N_5$.  On the other hand, we have seen that
by throwing
in tubes we can generate $|J_{23}-J_{45} | = \sqrt{N_1 N_5 N_p}$,
which can be much larger than  $2N_1 N_5$.  This mismatch can
presumably be accounted for by including the vectormultiplets in
the D1-D5 CFT, since angular momentum is also carried by the bosonic
components of these multiplets.

\subsec{Probing the BMPV black hole}

Some of the conclusions we have drawn in the previous section
actually rely on the probe analysis outside its domain of validity.
Since the probe, by definition,
should have a small effect on the ambient geometry, if we want to
ask questions about the maximal angular momenta we should really be
considering a supertube probing the BMPV black hole (or even better,
the as yet unknown generalization of BMPV mentioned above).  We
therefore now carry out this analysis. As we will see, the probe potential
will turn out to only depend on the harmonic function sourced by the
D4 branes,
and is independent on the D0 and F1 charges and the angular momenta.
This proves
that our previous inferences were in fact valid.

In the previous section we have also argued that by putting together tubes one cannot create an
overspinning BMPV black hole. The
probe analysis we now perform can also be used to show that one cannot overspin an already
existing underspinning BMPV black hole
by bringing in supertubes through the horizon.

As the first step, we need the lift of BMPV to 10 dimensions.  This was written down in
\HerdeiroAP\ and is reproduced in the
Appendix.  Next we T-dualize along the D1-branes, since this is more convenient for the probe
computation.  Using the T-duality
rules in \myers\ (see \MeessenQM\Buscher\ for earlier work) we obtain
\eqn\bmpvt{\eqalign{ \tg_{tt} &=  g_{tt} - g_{tz}^2/g_{zz} =   - H_5^{-1/2} H_1^{-1/2}
H_p^{-1}\cr \tg_{zz} &= 1/g_{zz} =
H_5^{1/2} H_1^{1/2} H_p^{-1}\cr \tB_{tz} &= -1 +  H_p^{-1} \cr \tB_{\phi_i z} &= J_i
H_p^{-1} }}
\eqn\bmptb{\eqalign{ \tg_{t \phi_i} &= g_{t \phi_i} -
g_{tz} g_{z \phi_i}/g_{zz}=g_{t \phi_i}  H_p^{-1} = - J_i H_5^{-1/2} H_1^{-1/2} H_p^{-1}
\cr
 e^{2 \tilde \phi} &= H_1^{3/2} H_5^{-1/2} H_p^{-1} \cr \tg_{\phi_i \phi_j} &= g_{\phi_i
\phi_j} - g_{z \phi_i} g_{z \phi_j}/g_{zz}
 \cr \tg_{\phi_1 \phi_1} &= H_5^{1/2} H_1^{1/2} r^2 \sin^2 \vt - J_1^2 H_5^{-1/2}
H_1^{-1/2} H_p^{-1} \cr
\tg_{\phi_2 \phi_2} &= H_5^{1/2} H_1^{1/2}  r^2 \cos^2 \vt - J_2^2
H_5^{-1/2} H_1^{-1/2} H_p^{-1} \cr \tg_{r
r} &= H_5^{1/2} H_1^{1/2}  \cr \tC^1_{t} &=  H_1^{-1} -1 \cr \tC^1_{\phi_i } &= J_i
H_1^{-1}\cr \tC^3_{t \phi_i z} &= - (C^2_{tz}
g_{\phi_i z} - C^2_{\phi_i z} g_{t z})/g_{zz} + C^2_{t \phi_i} = J_i H_p^{-1},}}
where $H_p, H_1$ and $H_5$ are the harmonic functions sourced by the F1 strings,
D0 and D4 branes respectively,
\eqn\extrat{
J_{1} =  {j \over 2 r^2} \sin^2 \vt, \ \ \ \ J_{2} =  {-j \over 2 r^2}
\cos^2 \vt, \ \ \
H_i= 1 + {R_i^2 \over r^2}~, }
and we only give the components of the forms
which we use in our calculations. This solution has a horizon at $r=0$.
The horizon
area is proportional to $\sqrt{ N_1 N_5 N_p - J^2}$, and matches the
entropy of the D1-D5-P system discussed in section 4. If $J^2
> N_1 N_5 N_p$ the solution has closed timelike curves. The solution
can be also be continued behind the horizon by
introducing the new coordinate $\rho^2 = R^2 + r^2 $.

We probe this metric with a D2 supertube carrying D0 and F1 charge.
On the worldvolume we turn on  $\cF_{tz}$ and
$\cF_{\theta z}$.   The former will eventually be set to $1$, but we
keep it arbitrary for now since we want to differentiate with respect
to it to get the F1 charge.  Note that $\theta$ is the worldvolume angular
coordinate, distinct from the coordinate $\vt$ appearing in the
supergravity solution. We allow $r$, $\vt$, and $\phi$ to vary arbitrarily as
 functions of $\theta$.  On the other hand, we take $t$ and $z$ to
coincide on the worldvolume and in spacetime.

The Wess-Zumino part of the brane action is
\eqn\wz{\eqalign{S_{WZ}&= T_2 \int \! \tC \wedge e^{\cF+\tB} \cr
 &= T_2\int\! dt \,dz \, d\theta
\left[ \Big(\tC_{t \phi z} + \tC_t \tB_{\phi z} -
\tC_{\phi} (\tB_{tz} + \cF_{tz})\Big) {\partial \phi \over \partial \theta}
+ \tC_t \cF_{\theta z}\right]
\cr
&= T_2 \int \! dt \,dz \, d\theta
\left[ j H_1^{-1} (1-\cF_{tz}) {\partial \phi \over \partial \theta}
+ \cF_{\theta z}
H_1^{-1} -\cF_{\theta z} \right].
}}
%

The Born Infeld part of the action is:
\eqn\bi{\eqalign{&S_{BI} = -T_2 \int\! dt\,dz\,d\theta\, e^{- \tilde \phi} \sqrt{- \det (\tg_{ab} +
\tB_{ab} + \cF_{ab})}\cr &
=-T_2 \int \! dt\,dz\,d\theta\, e^{- \tilde{\phi}} \ \cr &\times  \sqrt{-(\tg_{tt} \tg_{zz} \tg_{\theta
\theta} - \tg_{zz}\tg_{t
\theta}^2 - 2  \tg_{t \theta} (\tB_{tz} + \cF_{tz} ) (\tB_{\theta z} + \cF_{\theta z}) + \tg_{tt}
(\tB_{\theta z} + \cF_{\theta
z})^2 + \tg_{\theta \theta} (\tB_{tz} + \cF_{tz} )^2)} }}
%
The induced worldvolume  metric is
\eqn\bia{
\tg_{\theta \theta} = \tg_{\phi \phi} {\partial
\phi \over \partial \theta}{\partial \phi \over \partial \theta}  +
\tg_{rr} {\partial r \over \partial \theta} {\partial r \over \partial
\theta}}
%
%
The action then works out to be
\eqn\sqr{\eqalign{S_{BI} &= -T_2 \int\! dtdzd\theta\,
\Bigg[  (\cos^2 \vt  r^2 {\partial \phi \over \partial
\theta} {\partial \phi \over \partial \theta} +  {\partial r \over
\partial \theta} {\partial r \over \partial \theta}  )  H_5 H_1^{-1}
\Big(H_p^{-1} - H_p(H_p^{-1}  -1 + \cF_{tz})^2\Big) \cr
&\quad\quad\quad\quad\quad\quad\quad\quad  +H_1^{-2}   \Big(\cF_{\theta z}  + j
(1-\cF_{tz}) {\partial \phi \over \partial \theta}  \Big)^2 \Bigg]^{1/2}
}}
As usual, the BPS configuration is realized for $\cF_{tz}=1$.
 The total action then  simplifies to
\eqn\bi{S = S_{WZ}+S_{BI}= -T_2 \int \! dtdzd\theta\, \cF_{\theta z}=-
\int \! dtdz\, Q_0
}
and the Hamiltonian is simply
\eqn\h{H =\int\! dz \, ( Q_1 + Q_0)}
where $Q_1$ is the canonical momentum conjugate to $A_z$ as in \zc.    Thus, the configuration
is BPS for any shape, just as
expected.
 The shape of the tube is constrained by the explicit formula for $Q_1$,
which is
\eqn\qf{ Q_1 = {\partial L \over \partial \cF_{tz}}\big|_{\cF_{tz}=1}
= T_2 \int\! d\theta~ {H_5 \over \cF_{\theta z}}
(\cos^2 \vt r^2 {\partial \phi \over \partial
\theta} {\partial \phi \over \partial \theta} +  {\partial r \over
\partial \theta} {\partial r \over \partial \theta} ).}
The first thing to notice is that \qf\ is independent of the black
hole angular momentum. Indeed, it only depends on the induced
metric on the tube in the absence of $j$.  Furthermore, \qf\ only
depends on the harmonic function $H_5$, and so we could just as
well have been probing a pure D4-brane.  Thus, the previous simplified probe computation
captures the whole essence of the problem.

If we consider the simplest circular embedding $\phi=\theta$, with $r$ and
$\vt$ constant, we find
\eqn\rsq{ 2 \pi T_2 r^2 \cos^2 \vt ={ N_1 N_0\over H_5}}
which implies for the  D1-D5-P system:
\eqn\qq{
(N_1 N_p)_{\rm probe} =  2 \pi T_2r^2 \cos^2 \vt + (N_5)_{\rm hole} \, \cos^2 \vt.
}
We can bring the tube into $r=0$ as long as $(N_1 N_p)_{\rm probe} \leq (N_5)_{\rm hole} $.
If we bring the tube in at constant
$r$ and $\vt$ to ``crown'' the black hole, the tube will cross the horizon at angle $\cos^2 \vt =
{(N_1 N_p)_{\rm probe} \over
(N_5)_{\rm hole}} $.

Since the tube can be BPS for any radius one might think that there would exist configurations in
which the tube straddles the
horizon of the black hole, being partly inside and partly outside.  However, \qf\ implies that a
finite charge tube must have $
{\partial r \over \partial \theta} =0$ at the horizon, and so this cannot happen.

One can also use this probe computation to show that one cannot create closed timelike curves by
overspinning a regular BMPV black
hole. Let us take the charges of this black hole to be $N_p,N_1,N_5$ and its angular momenta to
be $J$, satisfying $J \leq
\sqrt{N_p N_1 N_5}$. We can only bring a tube with $\Delta J = \Delta N_p \, \Delta N_1$
inside the horizon if  $\Delta N_p \Delta
N_1 \leq N_5$. The resulting charges satisfy:
\eqn\overspin{\eqalign{(N_1+ \Delta N_1) (N_p +\Delta N_p) N_5 &= N_p N_1 N_5 +(N_p \,
\Delta N_1 + N_1\, \Delta N_p) N_5 + \Delta
N_p \, \Delta N_1 N_5 \cr &\geq J^2 + 2 N_5 \sqrt{ N_p \, \Delta N_1 N_1 \, \Delta N_p} +
N_5 \, \Delta J \cr &\geq J^2 + 2 J \,
\Delta J + (\Delta J)^2  = (J + \Delta J )^2}}
Thus, the resulting black hole is still underspinning, as expected
from chronology protection.

\newsec{ Conclusions and Outlook}

We have presented two ways to construct supersymmetric tubes with three charges and two or
three dipole moments. We then analyzed
the possibility that these configurations represent some of the hair of the spinning BMPV black
hole. We found that this
possibility passes some rather nontrivial tests. For example,
 the size of the tubes is always smaller than
the horizon circumference in the regime where both exist. Also, we showed how to use tubes to
construct a spinning three charge
black hole; the maximal angular momentum this black hole can carry is exactly the BMPV
maximal angular momentum.

Since the three charge supertubes can carry more angular momentum than a BMPV black hole
with the same charges, we have also
explored the possibility of using them to overspin the black hole. This would result in the creation
of closed timelike curves. We
have shown that this does not happen, providing a nice example of chronology protection.

As we have seen, the properties of the three charge supertubes
mesh nicely with the properties of black holes.   There are a number
of interesting directions to pursue.  By considering probes in the BMPV
background we gave an argument for the existence of supersymmetric
black solutions
which would generalize those of BMPV to unequal angular momenta, and possibly
to a tubular topology; it would be interesting to construct these.

It would of course
be very interesting to find the supergravity solutions for
arbitrary three charge supertubes, and to see if these can be put
in one-to-one correspondence with the states of the D1-D5-P system,
generalizing the work of \LuninJY\ - \MathurSV.  From the brane point of view,
it would be useful to be able to quantize the supertube configurations,
and give an account of the entropy in this description.
These are all problems to which we hope to return in the future.

Another BPS black hole in string theory is the four  dimensional
four charge black hole. Any three charges of this black hole can
be also carried by a U-dual of one of our three charge supertubes.
Hence, it would be also interesting to construct four charge
supertubes, and to see if they can account for the hair of this black hole.

\appendix{A}{10 dimensional lift of the BMPV black hole \HerdeiroAP}

The BMPV black hole lifts to the following 10 dimensional type IIB supergravity solution:
\eqn\bmpvo{\eqalign{ g_{tt} &=
H_5^{-1/2} H_1^{-1/2} (H_p-2)~\cr g_{zz} &=  H_5^{-1/2} H_1^{-1/2} H_p \cr g_{tz} & = -
H_5^{-1/2} H_1^{-1/2} (H_p-1)\cr g_{z
\phi_i} &= J_{i}  H_5^{-1/2} H_1^{-1/2}\cr g_{t \phi_i} &= - J_{i}  H_5^{-1/2} H_1^{-1/2}\cr
g_{\phi_1 \phi_1} &= H_5^{1/2}
H_1^{1/2}  r^2 \sin^2 \vt \cr g_{\phi_2 \phi_2} &= H_5^{1/2} H_1^{1/2}  r^2 \cos^2 \vt \cr
e^{2 \phi} &= H_1 H_5^{-1} g_s^2 \cr
C^2_{tz} &=  H_1^{-1} -1 \cr C^2_{\phi_i z} &= J_{i}  H_1^{-1}\cr C^2_{t \phi_i} &=  J_{i}
H_1^{-1} \cr C^2_{\phi_1 \phi_2} & =
(H_5 -1)r^2 \cos^2 \vt }}
where $i=1,2$,
\eqn\extrab{ J_{1} =  {j \over 2 r^2} \sin^2 \vt, \ \ \ \ J_{2} =  {-j \over 2 r^2} \cos^2 \vt, \ \ \
H_i= 1 + {R_i^2 \over r^2}~,
}
and $C^6$ is the same as $C^2$ with $H_1$ changed in $H_5$ and an extra 4-volume added.
This solution has a horizon at $r=0$. The
horizon area is proportional to $\sqrt{ N_1 N_5 N_p - J^2}$, and matches the entropy
of the D1-D5-P system discussed in section 4.
If $J^2 > N_1 N_5 N_p$ the solution has timelike curves.

\bigskip
\noindent {\bf Acknowledgements:} \medskip \noindent We thank Eric D'Hoker, Henriette
Elvang, Michael Gutperle, Gary Horowitz,
Clifford Johnson, Don Marolf, Radu Roiban, Masaki Shigemori, Wati Taylor, and especially
Samir Mathur for helpful discussions. The
work of IB and PK is supported in part by NSF grant 0099590.
 Any opinions, findings and conclusions expressed in this material
 are those of the authors and do not necessarily reflect the views
 of the National Science Foundation.

\listrefs
\end